\documentstyle[psfig,astrobib]{mn-ab}
\title[Ground-state OH observations towards NGC 6334]
{Ground-state OH observations towards NGC 6334}
\author[K. J. Brooks \& J. B. Whiteoak]
{K. J.~Brooks$^{1}$\thanks{Current institute: European Southern Observatory,
Casilla 19001, Santiago 19, Chile (kbrooks@eso.org)} and J. B.~Whiteoak$^{2}$\\
$^1$ Department of Physics, University of New South
Wales, Sydney 2052, NSW Australia\\
$^2$Australia Telescope National Facility, CSIRO, PO Box 76, Epping 2121, NSW Australia\\
}

\newcommand{\kms}{km\,s$^{-1}$}

\newcommand{\HII}{$\mathrm{H\,{\scriptstyle II}}$}

\begin{document}
\maketitle

\begin{abstract}

We have made observations of the four hyperfine transitions of the
$^{2}\Pi_{3/2}$, J=3/2 ground state of OH at 1612, 1665, 1667 and 1720 MHz
and the related 1.6-GHz continuum emission, towards NGC 6334 using the
Australia Telescope Compact Array. The observations covered all the major
radio continuum concentrations aligned along the axis of NGC 6334 (V, A to
F). We have detected seven OH masers plus a possible faint eighth maser;
two of these masers are located towards NGC 6334-A. Absorption at 1665 and
1667 MHz was detected towards almost all the continuum distribution. All
transitions showed non-LTE behaviour. The 1667-/1665-MHz intensity ratios
ranged from 1.0 to 1.2, significantly less than their LTE value of 1.8. The
results of the OH `Sum Rule' suggest that this discrepancy cannot be
explained solely by high optical depths. The 1612- and 1720-MHz
line-profiles showed conjugate behaviour whereby one line was in absorption
and the other in emission. In addition, the profiles commonly showed a flip
from absorption to emission and vice versa, which has been interpreted as a
density gradient. The OH line-to-continuum distribution, optical depth and
velocity trends are consistent with a bar-like shape for the molecular gas
which wraps around the continuum emission.

\end{abstract}


\section{Introduction} 
NGC 6334 is a prominent \HII\ region/molecular cloud complex located in the
Sagittarius arm of the Milky Way at a photometric distance of 1.7 kpc
\cite{Neckel78} ($1 {\mathrm pc} \approx 2$ arcmin). The complex contains
several recent and current star-forming sites which are embedded in an
elongated giant molecular cloud (GMC) extending over about 45 arcmin
(\citeNP{Dickel77}, \citeNP{Kraemer99}).

Far-infrared observations at wavelengths 80--250 microns by
\citeN{McBreen79} revealed six strong continuum concentrations, the peaks of
which are designated FIR-I to FIR-VI. Sources I to V are present along a
ridge of extended continuum emission which runs northeast--southwest,
parallel to the plane of our Galaxy, while the weak, extended source VI is
situated further south. Observations at 4.9 GHz by \citeN{Rodriguez82}
yielded six continuum concentrations also located along the
ridge. Designated A to F, some are coincident with the far-infrared
concentrations. The continuum ridge is coincident with a dark dust lane
seen at optical wavelengths (e.g \citeNP{Gardner75}). The dust lane appears
to be obscuring the optical ionized distribution, suggesting that it is
overlying the nebulosity. The radio continuum concentrations have been
individually observed in several projects (see discussion in Section
3.1). The results show that most of them have associated outflows, sites of
maser emission, or embedded protostars, all indicative of very active sites
of star formation.

To further investigate the physical conditions of the NGC 6334 complex, we
have carried out a series of observations of the 1612-, 1665-, 1667-, and
1720-MHz hyperfine transitions of the $^{2}\Pi_{3/2}$, J=3/2 ground state
of OH using the Australia Telescope Compact Array (ATCA)\footnote{The
Australia Telescope is funded by the Commonwealth of Australia for
operation as a National Facility managed by CSIRO.}.  These four
transitions are valuable in investigations of molecular clouds because they
can exhibit extended emission and absorption and maser emission.

A direct result of the statistical weights and transition probabilities of
the `main lines' (at 1665 and 1667 MHz) and `satellite lines' (at 1612 and 1720
MHz) is that, under conditions of local thermodynamic equilibrium (LTE),
their optical depths, $\tau_{1612}$, $\tau_{1665}$, $\tau_{1667}$,
$\tau_{1720}$, are in the ratio 1:5:9:1. Previous studies of OH have shown
that both the main-line and satellite-line intensities often show non-LTE
ratios.  The satellite lines can also exhibit a conjugate-type behaviour
whereby emission of one line is accompanied by absorption of the
other. This behaviour is explained in detail by \citeN{Elitzur92} and
references therein. The satellite-line behaviour can be used as a
diagnostic for the type of OH excitation process taking place in the region
and may also provide density constraints.  For example, radiative
excitation at high densities can produce 1612-MHz emission but collisional
excitation at low temperatures can lead to 1720-MHz emission
\cite{Elitzur92}.

The ratios of the optical depths of the four transitions give
\begin{equation}
\tau_{1612} + \tau_{1720} = \frac{\tau_{1665}}{5} + \frac{\tau_{1667}}{9}.
\end{equation}
This relationship is commonly termed the `Sum Rule'
(e.g. \citeNP{Rogers67}) and can be used to obtain optical depth measurements even in cases where LTE is not satisfied.  For low optical
depths and an extended continuum emission with a flat spectrum the Sum
Rule becomes
\begin{equation}
{T_{\rm 1612}} + {T_{\rm 1720}} = \frac{T_{\rm 1665}}{5} + \frac{T_{\rm
1667}}{9},
\end{equation}
where $T_{line}$ is the line temperature of the transition.

\section{Observations}

The ATCA observations were carried out with six different array
configurations between 1994 September and 1995 August. Details of the
instrument are given by \citeN{Frater92}. The antenna spacings ranged from
30 m to 6 km. All sets of observations were made with the correlator
configured to 2048 frequency channels over a bandwidth of 4 MHz, providing
a velocity resolution of 0.42 \kms.  Two orthogonal linear polarisations
were observed and then later averaged together. The observations were made
using standard procedures, cycling through frequencies centred on 1666,
1720 and 1612 MHz. The 1665- and 1667-MHz lines were both included in the
band centred at 1666 MHz.  Observations of PKS 1740-517 were used for phase
and bandpass calibration. Flux-density calibration was provided by
observing PKS 1934-638 (adopted as the ATCA primary calibrator), for which
flux densities of 14.3, 14.2, 14.0 Jy were adopted at 1612, 1666 and 1720
MHz respectively.

At 1.6 GHz, the half-power primary beamsize of the ATCA is 30 arcmin. For
observations centred on RA(B1950) = 17$^{h}$17$^{m}$10$^{s}$, Dec(B1950) =
$-35^{\circ}$49$\arcmin$00$\arcsec$, the half-power primary beam covered
all the major continuum concentrations along the ridge of NGC 6334.

The correlated spectral outputs from pairs of ATCA antennae were processed
using a package based on the Astronomical Image Processing System ({\sc
aips}) produced by the US National Radio Astronomy Observatory. The
continuum emission was subtracted from each spectral-line dataset and then
combined into a single continuum dataset. For all four spectral-line
datasets, the frequency scale was converted to a velocity scale with
respect to the local standard of rest (LSR). Amplitude and phase
self-calibration were applied to the 1665-, 1667- and 1720-MHz datasets,
all of which contained strong maser emission. Strong residuals remained in
the main-line channel maps ($\sim 50$ mJy beam$^{-1}$ for the images near
the central velocities of the masers) because the intensities of the two
masers both varied on time scales of a few months, producing different
values measured for each array-configuration.

Image analysis was carried out using {\sc miriad} \cite{Sault95} and {\sc
karma} \cite{Gooch96}.  Uniform and natural weighting were used for both
the continuum data and each channel of the line data.  Multi-frequency
synthesis \cite{Sault94} was also adopted for the continuum data. Each
image was deconvolved using the {\sc clean} algorithm of \citeN{Steer84} and
restored with diffraction-limited beams of $4.6 \times 7.2$ arcsec$^2$
(uniform weighting) and $25.8 \times 19.6$ arcsec$^2$ (natural
weighting). Each image was also corrected for the primary beam gain pattern
of the individual antennas. The uniform-weighted line images, with a
velocity resolution of 0.42 \kms, were used to investigate maser emission
and the natural-weighted line images, Hanning smoothed to a velocity
resolution of 0.7 \kms, were used to investigate the extended OH emission.

\section{Results}
\subsection{1.6-GHz Continuum distribution}

Fig. 1 shows the natural-weighted 1.6-GHz continuum image with an rms noise
level of 0.009 Jy beam$^{-1}$. The main emission concentrations are
labelled using the nomenclature of \shortciteN{Rodriguez82} and
\citeN{McBreen79} and their positions and fluxes are described in Table
1. The position of the far-infrared source NGC 6334-I(N) is marked with a
cross (see Section 3.1.6). The overall structure of NGC 6334-A to -F is
similar to that derived by \shortciteANP{Rodriguez82}
(\citeyearNP{Rodriguez82}, \citeyearNP{Rodriguez88}) at 4.9 GHz, and
\citeN{DePree95} at 8.3 GHz. Continuum emission at 1.6-GHz was also
detected towards NGC 6334-V and as a weak source south of NGC 6334-D (NGC
6334-D(S)). All of the concentrations are superimposed on faint
emission which extends northwest from the dark dust lane. This faint
emission, and also the two faint semi-circular emission regions near the
southern boundary of Fig. 1, follow the optical and infrared distributions.

\begin{figure*}
\psfig{file=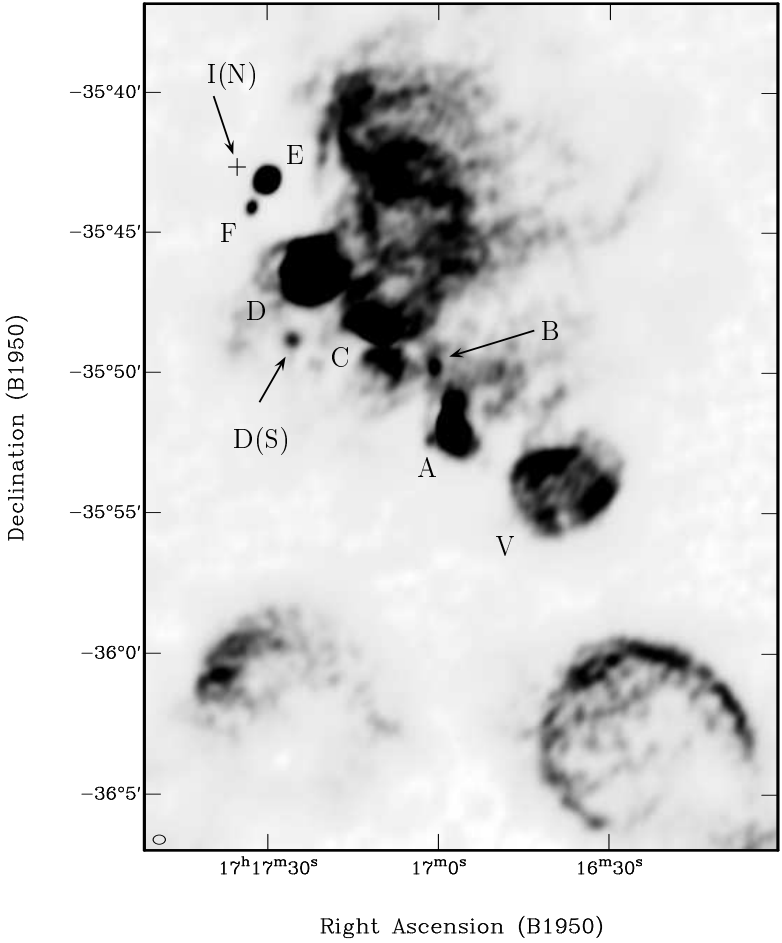,width=12cm,silent=}
\caption{NGC 6334: 1.6-GHz continuum emission imaged with a $25.8 \times 19.6$ arcsec$^2$ beam, shown bottom left}
\end{figure*}

\begin{table*}
\begin{minipage}{150mm}
\caption{1.6-GHz continuum parameters measured from Fig. 1. }
\begin{tabular}{@{}lccccc}
\hline
Region &	\multicolumn{2}{c}{Peak position\footnote{With a positional uncertainty of 1.5 arcsec}}		
        & Peak flux
		&Integrated flux &Velocity\\
		& RA(B1950)		& DEC(B1950)
	& (Jy beam$^{-1})$&(Jy)	& (\kms)\\ 
\hline 
NGC 6334-V	
		& 17$^h$16$^m$39\fs50	& $-35$\degr 53\arcmin 05\farcs9
	& $0.458 \pm 0.009$		& $11 \pm 2$  & \\
NGC 6334-A     
		& 	17 16 57.80 & $-35$ 51 43.9
        & $4.128 \pm 0.009$	& $12.9 \pm 0.7$  & $-0.1 \pm
        0.2$\footnote{Taken from H76$\alpha$ recombination line observations of \citeN{DePree95}}	\\ 
NGC 6334-B	
		& 	17 17 00.58	& $-35$ 49 47.2
	& $0.417 \pm 0.009$	& $0.417 \pm 0.009$	\\
NGC 6334-C	
		& 	17 17 10.98	& $-35$ 48 27.0
	& $2.295 \pm 0.009$	& $9.3 \pm 0.3$ & $-2.5
        \pm0.3$\footnote{Taken from H109$\alpha$ recombination line observations of \citeN{Reifenstein70}}	\\
NGC 6334-D     
		& 	17 17 22.87	& $-35$ 46 18.7 
        & $1.818 \pm 0.009$	& $23 \pm 2$	& $-2.5 \pm 0.3^c$\\ 
NGC 6334-D(S)     
		& 	17 17 25.52	& $-35$ 48 58.9
        & $0.266 \pm 0.009$	& $0.266 \pm 0.009$ \\ 
NGC 6334-E	
		& 17 17 29.61		& $-35$ 43 06.6
	& $1.636 \pm 0.009$	& $3.9 \pm 0.1$ \\
NGC 6334-F	
		& 17 17 32 53		& $-35$ 44 06.9
	& $0.433 \pm 0.009$	& $0.433 \pm 0.009$	& $-6.1 \pm 0.3^b$ \\
\hline
\end{tabular} 
\end{minipage}
\end{table*}

The uniform-weighted 1.6-GHz continuum image has an rms noise level of 0.02
Jy beam$^{-1}$. For this image, the main continuum concentrations listed in
Table 1 are discussed separately, with the exception of NGC 6334-D(S). This
source was not detected in the uniform-weighted image, which suggests it
may be a diffuse source of the order of 20 arcsec in diameter.

\subsubsection{NGC 6334-V}

This source was first detected by \shortciteN{McBreen79} and has a
supersonic bipolar outflow that is associated with shock-excited molecular
hydrogen gas (\citeNP{Fischer82}, \citeyearNP{Fischer85},
\citeNP{Straw892}, \citeNP{Kraemer99}). \citeN{Wolstencroft87} have
suggested that the outflow is powered by a protostar, designated IRS-24 by
\citeN{Straw89s}. This protostar may also be triggering nearby sites of
star formation.  \citeN{Kraemer95} have detected shock-excited NH$_{3}$
maser emission at the edge of the outflow. OH and H$_{2}$O maser emission
have also been detected (e.g. \citeNP{Forster89};
\citeNP{Caswell98}). \citeN{Jackson99} detected faint 8.485-GHz continuum
emission counterparts to FIR-V which may be additional protostars.

Fig. 2 reveals that the faint 1.6-GHz continuum emission in this region has
the morphology of a spherical shell with a diameter of about 3.5
arcmin. The shell is centred about 1 arcmin northeast of IRS-24 and parts
of it have recently been detected by \citeN{Jackson99} at 8.485 GHz. The
source of ionization is not known.

\begin{figure}
\psfig{file=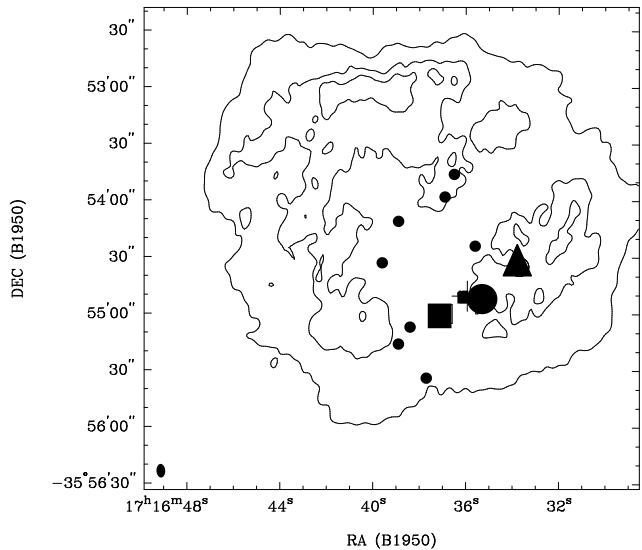,width=\columnwidth,silent=}
\caption{NGC 6334-V: 1.6-GHz continuum emission imaged with a $4.6 \times 7.2$ arcsec$^2$ beam, shown bottom left. The contour levels are 25,
40 and 50 mJy beam$^{-1}$. The other features shown and their positional
uncertainties are: {\it cross} -- OH maser ($\pm 0$\farcs1); {\it triangle}
-- NH$_{3}$ maser ($\leq \pm 3$ \arcsec); {\it small box} --
H$_{2}$O maser ($\pm 0$\farcs2); {\it large box } -- FIR-V ($\pm 30$\arcsec);
{\it small circle } -- star formation sites ($\pm 4$\arcsec); {\it large
circle} -- central protostar, IRS-24 ($\pm 4$\arcsec) .}
\end{figure}

\subsubsection{NGC 6334-A}

This component was first identified by \citeN{Schraml69}. A bipolar
morphology of the ionized gas was first detected by \citeN{Harvey83}. The
exciting star(s) of the region could be the infrared sources IRS-19 and/or
IRS-20 \cite{Straw89s}, identified as two B-type stars \cite{Harvey83}. The
elongated morphology and velocity gradients of the ionized gas were
interpreted as evidence of a confining molecular disk (\citeNP{Rodriguez88}
and \citeNP{DePree95}). Molecular-line observations by \citeN{Kraemer97}
revealed a flattened, massive, rotating structure of molecular gas which
extends 3 arcmin in the east--west direction (aligned perpendicular to the
continuum emission) and is centred near the IRS objects. In addition,
several small NH$_{3}$ emission clumps were detected within this
molecular structure.  One clump was found to be coincident with a site of
H$_{2}$O maser emission detected by \shortciteN{Rodriguez88} and another
with IRS-20. \citeN{Kraemer97} have suggested that these clumps are part of
a protocluster of stars which is condensing from the massive molecular
disk. \citeN{Jackson99} have detected an unresolved 8.485-GHz continuum
source at the location of IRS-20 which could be an optically thick \HII\
containing a star undergoing mass loss.

The 1.6-GHz continuum emission in Fig. 3 shows the same bipolar morphology
seen at other wavelengths, in which faint northern and southern extended
emission lobes extend out from a strong spherical shell. The lobes extend
over 3 arcmin north--south and the continuum shell has a diameter of 30
arcsec. The structure of the lobes is well defined, with the extended
emission narrowing at the northern edge of the central shell.

\begin{figure}
\psfig{file=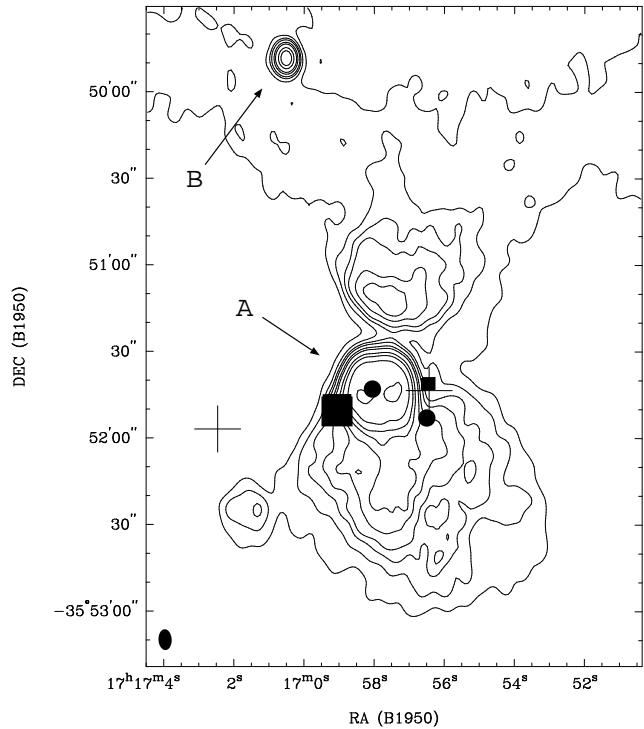,width=\columnwidth,silent=}
\caption{NGC 6334-A: 1.6-GHz continuum emission imaged with a $4.6 \times
7.2$ arcsec$^2$ beam, shown bottom left. The contour levels are 30, 40, 50,
60, 80, 100, 150, 200, 300 and 450 mJy beam$^{-1}$. The other sources shown
and their positional uncertainties are: {\it cross} -- OH masers
($\leq\pm 0$\farcs7); {\it small box} -- H$_{2}$O maser
($\pm 0$\farcs2); {\it large box} -- FIR-IV ($\pm 30$\arcsec); {\it circle} -- bright infrared sources, IRS-19 (left) and IRS-20 (right),
(both $\pm4\arcsec$). The figure also shows NGC 6334-B, the compact
northern source believed to be extragalactic -- see Section 3.1.3.}
\end{figure}

\subsubsection{NGC 6334-B}

Fig. 3 also shows the unresolved continuum emission of NGC 6334-B. The flux
of this source at 1.6 GHz was found to be 0.48 Jy, similar to the values
0.4, 0.6, 0.4 Jy at 4.8, 8.3 and 14.7 GHz respectively, derived by
\citeN{DePree95}. The source is superimposed on the faint, extended
emission seen northwards of NGC 6334-A. It is believed to be a
flat-spectrum extragalactic source and has been discussed by
\shortciteN{Rodriguez82} and \citeN{Moran90}. They found a variation of
intensity and angular size with wavelength which they explained in terms of
scattering by the ionized gas of NGC 6334-A.

\subsubsection{NGC 6334-C}

This \HII\ region was first discussed by \citeN{Schraml69}.
IRS-13 is believed to be responsible for the ionization of the region
\shortcite{Straw89s}. FIR-III appears to be centred to the southwest of
the main radio emission, even when its positional uncertainty of $\pm30$
arcsec is considered. Because of the relative diffuseness of the radio
emission and lack of maser sources NGC 6334-C was thought to be one of the
older areas of NGC 6334. However, \citeN{Straw891} have detected a
significant number of low-to-intermediate-mass pre-main-sequence objects in
this region. \citeN{Persi82} have detected a compact infrared source as
well as Br$\alpha$ line emission towards an infrared peak that may be
associated with an extreme young object.

The 1.6-GHz continuum emission (Fig. 4) is extended to the north but the
southern edge is clearly compressed, in agreement with the observations by
\citeN{DePree95}.

\begin{figure}
\psfig{file=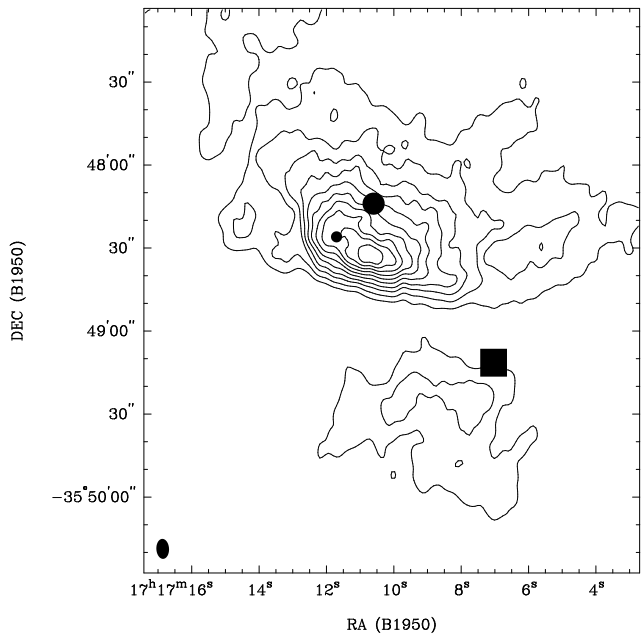,width=\columnwidth,silent=}
\caption{NGC 6334-C: 1.6-GHz continuum emission distribution imaged with a
$4.6 \times 7.2$ arcsec$^2$ beam, shown bottom-left. The contour levels are
45, 60, 80, 100, 125, 150, 175, 200, 225 and 250 mJy beam$^{-1}$. The other
sources shown and their positional uncertainties are: {\it large box} --
FIR-III ($\pm30\arcsec$); {\it large circle} -- compact infrared
source ($\pm10$\arcsec); {\it small circle} -- Br$\alpha$ line
emission ($\pm10$\arcsec).}
\end{figure}

\subsubsection{NGC 6334-D}

This is another of the three \HII\ components first discussed by
\citeN{Schraml69}. The presence of an intense arc of H$_{2}$ emission lying
to the southeast of FIR-II led \citeN{Straw892} to suggest that this region
of the molecular cloud has been swept relatively clear of gas and dust by
an outflow responsible for the H$_{2}$ emission. As yet, no further
observations have been made to confirm the presence of this outflow. Two
possible candidates that may be powering the outflow are IRS-23, an early
B-type star, and IRS-24, an O-type star thought to be also responsible for
the continuum emission \shortcite{Straw89s}. Because of its extended,
spherical continuum structure NGC 6334-D is thought to be one of the more
evolved \HII\ regions, although the region contains, in addition to the
infrared objects already mentioned, a significant number of
pre-main-sequence objects \cite{Straw891} and H$_{2}$O maser emission
\cite{Moran80}.

Consistent with previous observations at higher frequencies, the 1.6-GHz
continuum emission (Fig. 5) is spherical and featureless with a diameter of
2.5 arcmin. The arc of intense H$_{2}$ emission is along the south-eastern
edge.

\begin{figure}
\psfig{file=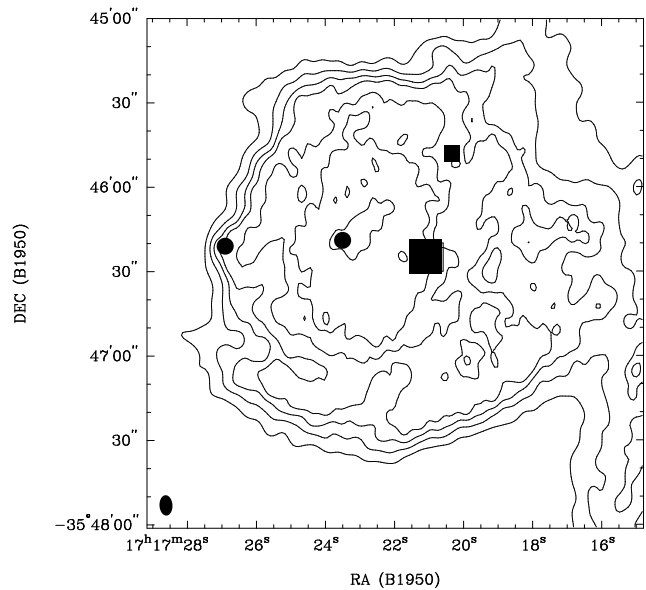,width=\columnwidth,silent=}
\caption{NGC 6334-D: 1.6-GHz continuum emission distribution imaged with a
$4.6 \times 7.2$ arcsec$^2$ beam, shown bottom left. The contour levels are
30, 40, 50, 60, 80, 100 and 150 mJy beam$^{-1}$. The other sources shown
and their positional uncertainties are: {\it small box} --
H$_{2}$O maser ($\pm20\arcsec$); {\it large box} -- FIR-II
($\pm 30\arcsec$); {\it circle} -- IRS-23 (left), IRS-24
(right), (all $\pm4\arcsec$).}
\end{figure}

\subsubsection{NGC 6334-E}

NGC 6334-E was first detected by \shortciteN{Rodriguez82} as a spherical
and featureless radio-continuum source. A cluster of B-type stars is
thought to be responsible for the ionization \cite{Tapia96}. To the
northwest of it is the well-studied source NGC 6334-I(N). Its position is
shown on Fig. 1. This source is the brightest 1-mm continuum source
observed in NGC 6334 \cite{Cheung78}. It is likely to be an active
star-formation region in the early stages of evolution (see
\shortciteNP{Tapia96} and references therein).

The 1.6-GHz continuum emission of NGC 6334-E (Fig. 6) is consistent with an
evolved spherical shell of diameter 40 arcsec.  No continuum emission at
1.6 GHz was detected towards NGC 6334-I(N) to an upper limit of 9 mJy
beam$^{-1}$.

\begin{figure}
\psfig{file=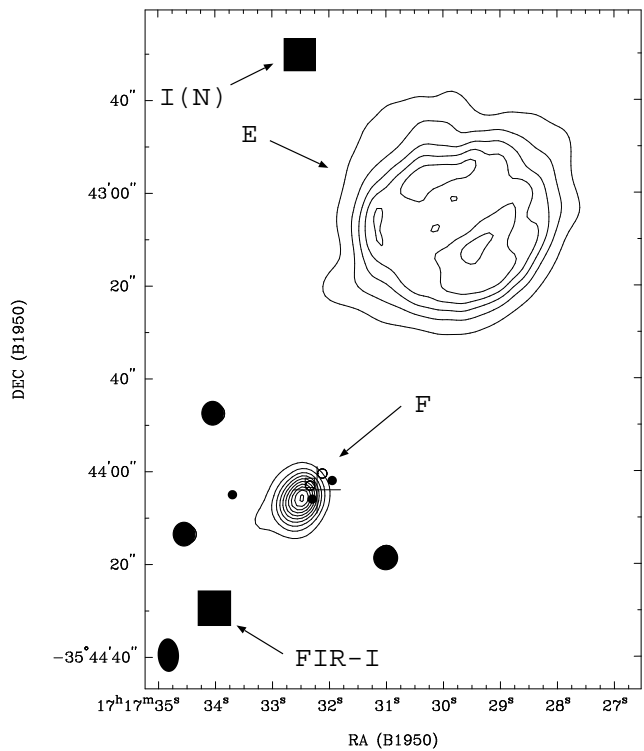,width=\columnwidth,silent=}
\caption{NGC 6334-E, -F: 1.6-GHz continuum emission distribution imaged
with a $4.6 \times 7.2$ arcsec$^2$ beam, shown bottom left. The contour
levels are 25, 50, 75, 100, 125, 150, 175, 200, 225, 250 mJy
beam$^{-1}$. The other sources shown and their positional uncertainties
are: {\it cross} -- OH masers ($\pm0$\farcs1); {\it large circle} --
NH$_{3}$ masers ($\leq\pm3$\arcsec); {\it open small circle} -- CH$_{3}$OH
masers ($\pm0$\farcs2); {\it open small box } -- H$_{2}$O maser
($\pm0$\farcs4); {\it large box} -- FIR-I ($\pm30\arcsec$), I(N)
($\pm0$\farcs5); {\it small circle } -- IRS-3(left), IRS-1 (middle), IRS-2
(right), (all $\pm\leq7$\arcsec). The H$_{2}$O maser and one CH$_{3}$OH
maser are coincident.  }
\end{figure}

\subsubsection{NGC 6334-F}

NGC 6334-F is an ultra-compact \HII\ region that is optically obscured. It
has been described as a `nozzle' by
\shortciteN{Rodriguez82}. High-resolution continuum observations by
\citeN{Ellingsen96} and \citeN{DePree95} show evidence of compression on
the north-eastern edge. FIR-I is located 30 arcsec to the southeast but in
view of its positional uncertainty of 30 arcsec, may in reality be
associated with the object. The excitation of the \HII\ region is caused by
IRS-1, most likely a massive star \cite{Harvey83}. From recent thermal dust
imaging by \citeN{Persi98} it appears that the circumstellar dust of IRS-1
is intermixed with the ionized gas associated with the \HII\ region. A
molecular bipolar outflow extends across the \HII\ region in a
northeast--southwest direction \citeN{Bachiller90}. Sites of NH$_{3}$
maser emission \cite{Kraemer95} and shocked H$_{2}$ emission \cite{Persi96}
occur at the ends of the outflow.  A velocity gradient is also detected in
the ionized gas \cite{DePree95}. It is still not clear if the outflow is
powered by IRS-1 or a nearby infrared source, IRS-2 \cite{Harvey83}. Maser
sites for transitions of OH (e.g. \citeNP{Gaume87}; \citeNP{Caswell98}),
H$_{2}$O (e.g. \citeNP{Forster89}) and CH$_{3}$OH (e.g. \citeNP{Caswell97})
are located towards or near the \HII\ region.

The 1.6-GHz continuum emission is barely resolved in Fig. 6, revealing a
slight extension in the southeast, northwest direction. The measured flux
at 1.6 GHz was 0.43 Jy, consistent with the values of 0.59 measured at 8.5
GHz \shortciteN{Ellingsen96} and 0.4 and 0.1 Jy measured at 14.5 and 4.8
GHz respectively \cite{DePree95}.

\subsection{OH maser emission}

Seven OH masers were unambiguously detected. There is strong observational
evidence to suggest that OH masers are linked to massive young stars and
ultracompact \HII\ regions even though the details of their pumping
mechanism remain unclear (e.g. \citeNP{Elitzur92}). All of the masers
detected are located along the main continuum axis and are associated with
three of the components discussed earlier. Their presence suggests that
star formation is still occurring in these regions. Table 2 lists the peak
intensity, velocity and position for each maser obtained from a
2-dimensional Gaussian fit to the peak emission. Fig. 7 shows the seven
individual spectra.

The strongest OH masers are associated with NGC 6334-V and NGC 6334-F and
are well documented from previous studies. We have obtained positional
uncertainties of $\pm0.1$ arcsec. For NGC 6334-V our 1665- and 1667-MHz
maser positions differ by 0.7 arcsec and their positions are coincident to
within 0.3 arcsec of the previous detections by \citeN{Caswell98} which
have a positional error of $\pm0.4$ arcsec. It appears that the location of
the two masers differ slightly. In contrast, for NGC 6334-F our 1665 and
1667 masers are coincident in position to within their positional errors
and to within 0.3 arcsec of the respective positions measured by
\citeN{Caswell98}. The 1720-MHz maser is offset in position from the
main-line masers by 0.8 arcsec. Its position differs by 0.4 arcsec from the
previous detection by \citeN{Gaume87} which has a
positional error of $\pm0.15$ arcsec. A possible new detection of a faint
1612-MHz maser associated with NGC 6334-F is discussed in Section 3.3.3.

Two 1665-MHz masers were detected towards NGC 6334-A. In contrast to the
other 1665-MHz masers which are bright and peaked near $-8.7$ \kms, these two
masers are faint with single features at more positive velocities.  One
has a velocity of $-0.8$ km s$^{-1}$ and occurs at the western edge of the
continuum peak (see Fig. 3). It is offset from the nearby H$_{2}$O maser by
2.7 ($\pm0.3$) arcsec. The other OH maser is situated 1 arcmin east of the
continuum shell and has a velocity of $-0.3$ km s$^{-1}$. It does not
appear to be associated with any other detected features. The flattened
molecular structure described in Section 3.1.2 extends out past this new
maser site. If star formation is thought to be occurring within this
structure then the maser may be pinpointing a new site of star formation.

\begin{table*}
\begin{minipage}{150mm}
\caption{Parameters of the peak OH maser emission.} 
\begin{tabular}{@{}lcccccc}
\hline
Region  	&OH line	& \multicolumn{2}{c}{Peak position}	 
			
        & Peak Intensity	& Velocity \\
	
		& (MHz) 	& RA(B1950) & DEC(B1950)
			
	& (Jy)\			&(km s$^{-1}$) \\ 
\hline 
NGC 6334-V     & 1665  & 17$^h$16$^m$35\fs94 $\pm0.08$    &
$-35$\degr54\arcmin51\farcs2 $\pm 0.1$
			
	& $53.8 \pm 0.9$	& $-8.622 \pm 0.007$ 	\\ 
                & 1667  & 17 16 $35.94 \pm 0.01$    & $-35$ 54 $50.5 \pm 0.1$
			
        & $72 \pm 1$	& $-8.7 \pm 0.7$	\\
\\ 
NGC 6334-A     & 1665  & 17 16 $56.43 \pm 0.03$    & $-35$ 51 $43.6 \pm 0.3$
			
        & $0.24 \pm 1$ 	& $-0.8 \pm 1$  \\ 
        	& 1665  & 17 17 $02.46 \pm 0.02$    & $-35$ 51 $56.9 \pm 0.7$
			
        & $0.14 \pm 1$	& $-0.3 \pm 2$\\
\\

NGC 6334-F	& 1665 	& 17 17 $32.22 \pm 0.01$	& $-35$ 44 $04.0
\pm 0.1$
			
	& $120.1 \pm 0.9$	& $-8.860  \pm 0.003$  \\
		& 1667  & 17 17 $32.23 \pm 0.01$    & $-35$ 44 $04.1 \pm 0.1$
			
        & $28.1 \pm 0.5$ & $-8.94 \pm 0.03$  \\
		& 1720 	& 17 17 $32.23 \pm 0.01$    & $-35$ 44 $04.8 \pm 0.1$
			
	& $27.1 \pm 0.7 $	& $-10.16 \pm 0.02$  \\
\hline 
\end{tabular} 
\end{minipage}
\end{table*}

\begin{figure}
\psfig{file=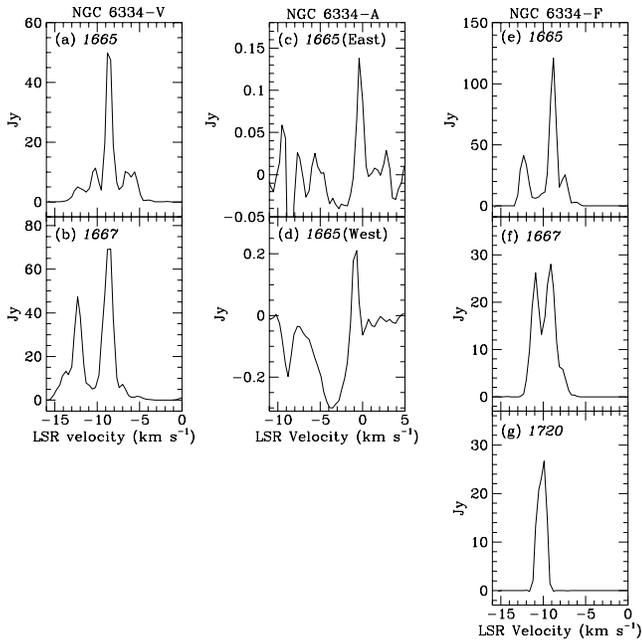,width=\columnwidth,silent=}
\caption{Line profiles of OH maser emission detected towards NGC 6334-V (a,
b), NGC 6334-A (c, d) and NGC 6334-F (e--g). The frequencies are labelled
in MHz.}
\end{figure}

\subsection{Extended OH emission and absorption}

The OH absorption profiles (see e.g. Fig. 9) reveal two molecular clouds
located along the line of sight to NGC 6334, one with velocities extending
from $-15$ to $+2$ \kms, the other with a well-defined velocity near $+6$
\kms. The more negative velocity absorption is confined towards NGC 6334
whereas the narrow absorption at $+6$ \kms\ is also detected towards other
nearby \HII\ regions (e.g. \citeNP{Whiteoak74}) and is believed to
originate in an extended overlying foreground cloud.

\subsubsection{Distribution of extended 1667-MHz OH absorption}

Except at high optical depths, the line-to-continuum ratio,
$T_{1667}/T_{\rm c}$, is related to OH column densities, and can be used to
investigate the large-scale spatial distribution of the OH cloud across
NGC 6334. However, the interpretation of the results can be confused by the
presence of either maser emission or continuum emission located in front of
the OH gas.

Fig. 8. shows the peak $T_{1667}/T_{\rm c}$ distribution superimposed on
continuum emission brighter than 45 mJy beam$^{-1}$ (five times the rms
noise level). Strong maser emission towards NGC 6334-F and NGC 6334-V
inhibits meaningful values for these regions. The OH cloud is concentrated
towards NGC 6334-A and -E and it appears to be offset to the northwest of
NGC 6334-C and, to a lesser extent, NGC 6334-D. Although the results
suggest that the OH column density in front of these two continuum
components is low, it may be that the bulk of the cloud extends behind
them.

\begin{figure}
\psfig{file=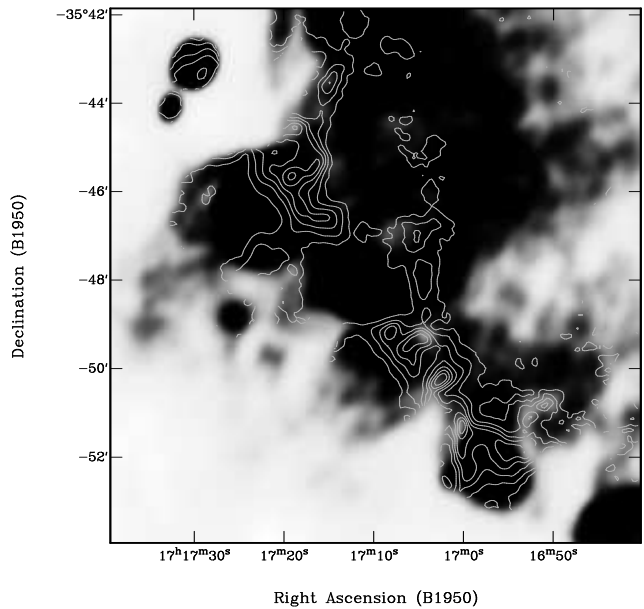,width=\columnwidth,silent=}\caption{$T_{1667}/T_{\rm
c}$ contours overlaid on the 1.6-GHz continuum emission both derived with a
$25.8 \times 19.6$ arcsec$^2$ beam. The $T_{1667}/T_{\rm c}$ is averaged
over $-4.0$ to $-1.9$ \kms. The contour levels are $-0.88$, $-0.75$,
$-0.63$, $-0.50$, $-0.375$, $-0.25$, $-0.125$. The grey-scale continuum has
been cut off at 45 mJy beam$^{-1}$.}
\end{figure}

The south-eastern edge of this OH cloud coincides with a sharp edge of the
dark dust lane mentioned previously, supporting a mixing of gas and
dust. High line-to-continuum values occur in localised regions of
approximately 30 arcsec in diameter within the OH concentration which
extends northeast of NGC 6334-A and southwest of NGC 6334-C. The highest
line-to-continuum value of $-0.9$ occurs at a velocity of $-2.6$ \kms\
towards RA(B1950) = 17$^{h}$17$^{m}$02\fs5, Dec(B1950) = $-35^{\circ} 50
\arcmin 14 \arcsec$. The compression of the radio continuum emission at the
southern edge of NGC 6334-C (see Fig. 4) may be a result of an interaction
with the OH cloud.

\subsubsection{Non-LTE behaviour and the OH Sum Rule}

Fig. 9 shows the line profiles towards NGC 6334-A, -E and -D at the
positions that are listed in Table 3. The rms noise level for each
transition is less than 0.01 Jy beam$^{-1}$.  Also shown are the profiles of
the weighted sums $\sum_{\rm s}$ (solid) and $\sum_{\rm m}$ (dashed),
defined by Eq. 2. The profiles for these three sources have been selected
because they are not affected by the strong maser emission associated with
the other sources and they show the typical behaviour seen for the OH
cloud.  

\begin{figure}
\psfig{file=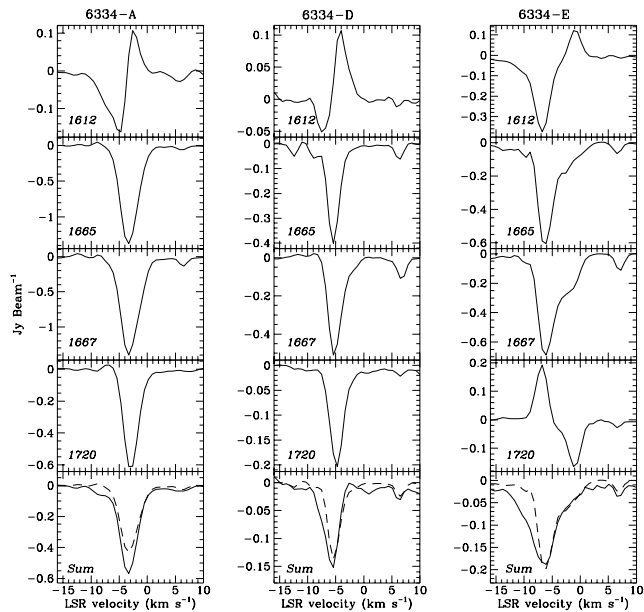,width=\columnwidth,silent=}
\caption{Line profiles of OH thermal emission and absorption towards
NGC 6334-A, -D and -E. The weighted sums of the satellite lines,
$\sum_{\rm s}$ (solid) and the main lines, $\sum_{\rm m}$ (dashed)
are also shown.  }
\label{Lines2}
\end{figure}

The satellite-line profiles show specific examples of the conjugate
behaviour mentioned in Section 1.  In addition, some profiles exhibit a
`flip' from emission to absorption with a change in velocity.  This flip,
first detected in an individual feature by \citeN{Caswell74} has been
discussed in some detail in recent studies (e.g. \citeNP{Frayer98};
\citeNP{Langevelde95}). An interpretation is that the more
negative-velocity gas, which shows 1612-MHz absorption, has OH column
densities such that {\it N}$_{OH}/\Delta {\mathrm V} \leq 10^{15}$
cm$^{-2}$ km$^{-1}$ s where $\Delta$V is the full-width at half maximum of
the absorption profile.  For the more positive-velocity gas, which shows
1612-MHz emission, the OH column density is given by {\it N}$_{OH}/\Delta
{\mathrm V} \geq 10^{15}$ cm$^{-2}$ km$^{-1}$ s. The velocity at which this
flip occurs represents the transitional density, {\it N}$_{OH}/ \Delta
{\mathrm V} \approx 10^{15}$ cm$^{-2}$ km$^{-1}$ s. For the OH main-line
absorption profiles in NGC 6334, $\Delta{\mathrm V} \approx 3$ \kms. Using
the constant OH/H$_{2}$ abundance ratio of 10$^{-7}$ \cite{Langer89}, the
corresponding transitional H$_{2}$ column density is $3 \times 10^{22}$
cm$^{-2}$.

Table 3 lists values of the peak line-to-continuum ratio measured at
specific positions for all four transitions. The rms noise level in both
the continuum and line images introduced an uncertainty for each
measurement that was below 10 per cent. In cases where the profiles `flip'
from emission to absorption, the peak with the highest absolute value was
used. The velocities taken from the 1665-MHz absorption and 1612-MHz
emission are most positive ($-2.2$ \kms) just south of NGC 6334-C, become
more negative towards both ends of the continuum ridge, and reach values of
$-6.1$ \kms\ towards NGC 6334-V and $-8.2$ \kms\ towards NGC 6334-F. This
velocity trend is supported by the velocity trends of the OH masers (see
Table 2) and has been detected in observations of other molecules
(e.g. \citeNP{Kraemer99}). The trend is also present in the limited \HII\
velocities available.

\begin{table*}
\begin{minipage}{\textwidth}
\caption{Parameters derived from the absorption and emission profiles
at selected positions within NGC 6334.} 
\begin{tabular}{@{}ccccccccccc}
\hline
Region &\multicolumn{2}{c}{Position}  &Velocity &$\frac{T_{1612}}{T_{c}}$
	&$\frac{T_{1665}}{T_{c}}$ &$\frac{T_{1667}}{T_{c}}$
	&$\frac{T_{1720}}{T_{c}}$ & $\frac{T_{1667}}{T_{1665}}$ &
	$\sum_{\rm m}$\footnote{Values of the weighted sums defined by
	Eq. 2 to within an error of $\pm0.02$} & $\sum_{\rm s}$ $^a$\\

 & RA(B1950) 	& DEC(B1950)	&(\kms) & & & & & 
	& \multicolumn{2}{c}{(Jy beam$^{-1}$)} \\
	
\hline 
NGC 6334-V	&17$^h$16$^m$35\fs48	&-$35$\degr55\arcmin17\farcs2
&$-6.1$\footnote{Velocity of $T_{1612}$ intensity peak to within an error of
$\pm 0.4$ \kms. No main-line extended emission or absorption measurements\\
can be made towards NGC 6334-V and -F because of the strong maser emission
at these transitions.} &$+ 1.5$	&---
&--- &$-0.7$ &---\\   

NGC 6334-A	&17 16 58.33	&$-35$ 51 38.0	&$-3.3$\footnote{Velocity of
$T_{1665}$ intensity peak to within an error $\pm 0.4$ \kms}&$- 0.1$ 
&$-0.84$  &$-0.83$ 	&$-0.40$ 	& $0.99 \pm 0.03$	 & 0.44  &
0.58  \\

NGC 6334-C	&17 17 09.75	&$-35$ 49 06.0	&$-2.1^{\it c}$ & $-0.17$ 
& $-0.36$ 	& $-0.36$ 	& $< \pm  0.02$		& $1.0 \pm 0.4$	 \\

NGC 6334-D	&17 17 22.16	&$-35$ 46 06.6	& $-5.4^{\it c}$ & $+0.09$  &
$-0.30$ 	& $-0.37$ &$-0.14$ 	& $1.22 \pm 0.09$		& 0.19
& 0.18  \\

NGC 6334-E	&17 17 29.13	&$-35$ 43 09.7	&$-6.1^{\it c}$ &$-0.28$ 
& $-0.48$	& $-0.47$  &$-0.18$ 	& $0.98 \pm 0.06$	 & 0.17 	&
0.15 \\

NGC 6334-F	&17 17 32.32	&$-35$ 44 03.9	&$-8.2^{\it b}$ &$+0.55$
&--- 	& --- &---	& ---	\\

\hline
\end{tabular} 
\end{minipage}
\end{table*}

For NGC 6334-D and -E, Eq. 2 is approximately satisfied suggesting that the
OH optical depths towards these regions are low. Therefore the 1665- and
1667-MHz line-to-continuum values of these two regions are valid
approximations of the actual optical depths at each transition. For NGC
6334-A the sum of the line-to-continuum ratios for the satellite lines
($0.58 \pm 0.02$) is higher than the weighted sum for the main lines ($0.44
\pm 0.02$). This discrepancy is consistent with higher optical depths and
therefore the line-to-continuum measurements as approximations of optical
depth should be used conservatively.

In all measurable cases, the main-line profiles also show non-LTE line
ratios. The $T_{1667}$/$T_{1665}$ ratios range from 1.0 to 1.2. These
values differ significantly from the LTE value of 1.8. Since the we know
from the results of Eq. 2 that the optical depths are low towards NGC
6334-D and -E, the non-LTE ratios for these two regions can only be
explained in terms of high optical depth if the cloud is clumping on
smaller scales.

\subsubsection{Use of the satellite lines to probe individual OH cloud
cores}

Figs 10 to 12 show the velocity maps of the 1612-MHz emission and
absorption for NGC 6334-V, -A, and -D. The velocity increment of each image
is 0.7 \kms. Only the 1612-MHz transition is shown because the
corresponding 1720-MHz transition has a similar distribution and does not
demonstrate the flip from emission to absorption as clearly as the 1612-MHz
transition. In all cases there is significant 1612-MHz emission which
presumably results from radiative excitation of dense OH (H$_2$ column
densities $\geq 5 \times 10^{22}$ cm$^{-2}$) by nearby infrared objects.

\begin{description}
\item {\it NGC 6334-V:} The set of velocity maps (Fig. 10) show 1612-MHz
emission present over a velocity range of $-11$ \kms\ to $-2$ \kms. The
distribution does not mirror the continuum emission shown in Fig. 2, but
consists of a concentration which extends over 2.5 arcmin in the
north--south direction and 2 arcmin east--west, covering the region of
enhanced infrared emission as might be expected. The distribution coincides
with a region of CO(2--1) emission \cite{Kraemer98}. The possible sites of
triggered star formation detected by \citeN{Straw89s} trace an arc around
the eastern edge of the 1612-MHz emission. The change of this emission with
velocity suggests an expanding filled-shell structure. At some velocities a
relatively sharp decrease in intensity is present at the south-western edge
of the peak emission. The 1720-MHz OH transition is in absorption over the
same velocity range as the 1612-MHz transition, and has the same
distribution. To the west of the main 1612-MHz emission is an area of faint
absorption. It is associated with faint 1720-MHz emission, and here, away
from the infrared emission, collisional excitation may be responsible.

\begin{figure}
\psfig{file=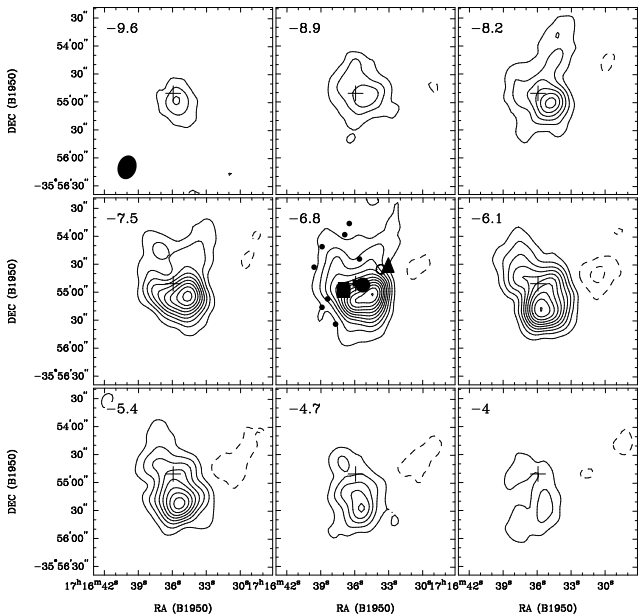,angle=270,width=\columnwidth}
\caption{NGC 6334-V: 1612-GHz emission (bold) and absorption (dashed)
distribution imaged with a $25.8 \times 19.6$ arcsec$^2$ beam, shown bottom
left. The contour levels are $-50$, $-25$, 25, 50, 75, 100, 125, 150, 175,
200, 225, 250 and 275 mJy beam$^{-1}$. The sites of OH masers ({\it cross})
are shown on all images. The other symbols are described in the caption of
Fig. 2}
\end{figure}

Although the extended main-line OH absorption is masked by two bright
masers (see Fig. 7), a comparison of peak line-to-continuum ratios at 1612
MHz and 1720 MHz, listed in Table 3, indicates anomalous behaviour. The
average of the two values ($+0.4$) is positive, whereas for a simple
OH-absorption environment the Sum Rule decrees that the value must be
negative. One explanation might be that radiative excitation of a dense OH
cloud by the embedded infrared objects has resulted in low-gain maser
amplification at 1612-MHz.

\item {\it NGC 6334-A:} The profiles in Fig. 9 show, 1612-MHz
absorption between $-8$ and $-6$ \kms\ and then 1612-MHz emission to $+1$
\kms. At 1720 MHz, emission between $-10$ and $-5$ \kms\ is followed by
deep absorption extending to $+1$ \kms. The weighted-sum profile has a
shape similar to the main-line profiles, as predicted by the Sum
Rule. Conjugate satellite-line behaviour is obviously present and the
extent is given by the deviations of the individual spectra from the
weighted-sum profile. Accordingly, for the 1612-MHz transition, the
deviations would show a profile similar in shape to the 1612-MHz emission
profile, but the absorption intensity would be reduced and emission
intensity increased (to about 0.2 Jy beam$^{-1}$). For the 1720-MHz
transition the difference spectrum would show exact conjugate behaviour.

The set of velocity maps (Fig. 11) show that the overall distribution of
the 1612-MHz absorption and emission is spherical, with an average diameter
of about 1 arcmin, and coincides with the central region of bright
continuum emission in Fig. 3. The extent is similar to that for the NH$_3$
emission \cite{Kraemer97}. At a velocity of $-4$ \kms\ the distribution
splits into absorption and emission components, diametrically offset. At
the extreme velocities, the relative offset virtually disappears, ruling
out rotation. The 1720-MHz transition exhibits conjugate behaviour with the
same distribution as the 1612-MHz transition, including the splitting at
$-4$ \kms.

The results are consistent with satellite-line anomalies resulting from OH
excitation by the infrared objects embedded in the cloud and a density
gradient that increases with the more positive observed velocities of the
molecular gas. This density gradient may also explain the distribution at a
velocity of $-4$ \kms\ and is discussed further in Section 4.2.

\begin{figure}
\psfig{file=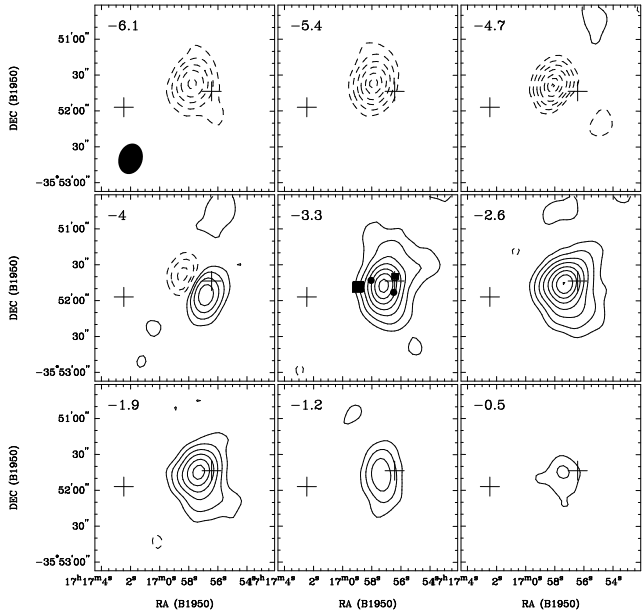,angle=270,width=\columnwidth,silent=}
\caption{NGC 6334-A: 1612-GHz emission (bold) and absorption (dashed)
distribution imaged with a $25.8 \times 19.6$ arcsec$^2$ beam, shown bottom
left. The contour levels are $-235$, $-210$, $-175$, $-125$, $-85$, $-50$,
$-25$, 25, 50, 85, 125, 175, 225, 275, 325 mJy beam$^{-1}$.  The sites of
the OH masers ({\it cross}) are shown on all images. The other symbols are
described in the caption of Fig. 3.}
\end{figure}

\item {\it NGC 6334-D:} For the 1612-MHz transition, the profiles
in Fig. 9 show absorption from $-10$ \kms\ to $-5$ \kms\ then emission
extending to 0 \kms. The 1720-MHz transition is totally in
absorption. Comparison of these profiles with the weighted-sum profile
suggests that anomalous satellite-line behaviour is present only at
velocities more positive than $-7$ \kms. Here, the enhanced 1612-MHz emission
is accompanied by enhanced 1720-MHz absorption.

The set of velocity maps (Fig. 12) reveal that the distribution of the
1612-MHz absorption and emission is not completely coincident with the
region of brightest continuum emission. Instead of an approximately
spherical distribution centred near IRS-24, 1612-MHz emission is
concentrated along an arc to the west of the object. The 1720-MHz
absorption was found to have a similar distribution. The offset
locations are consistent with the extended OH cloud distribution traced by
the 1667-MHz absorption in Fig. 8. There are no velocity trends to confirm
the presence of a possible outflow in the region. The anomalous
satellite-line behaviour is presumably due to the OH excitation by one or
more of the infrared sources. It may be significant that for both this and
the previous case, the \HII\ velocities fall within the velocity range of
the 1612-MHz emission.

\begin{figure}
\psfig{file=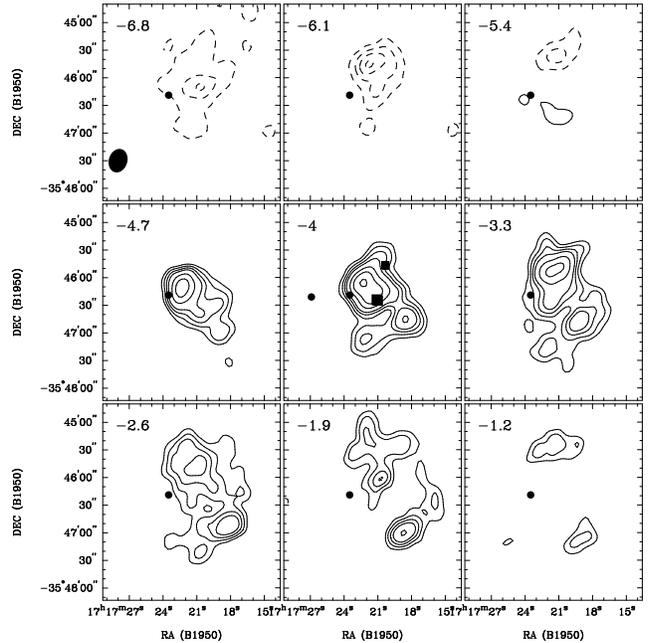,angle=270,width=\columnwidth,silent=}
\caption{NGC 6334-D: 1612-GHz emission (bold) and absorption (dashed)
distribution imaged with a $25.8 \times 19.6$ arcsec$^2$ beam, shown bottom
left. The contour levels are $-100$, $-75$, $-50$, $-25$, 30, 40, 50, 60,
80, 100, 125, 150, 175, 200, 225, 250 and 275 mJy beam$^{-1}$. IRS-24 ({\it
small circle}) is shown on all images. The other symbols are described in
the caption of Fig. 5.}
\end{figure}

\item {\it NGC 6334-E:} Anomalous satellite behaviour dominates the
1612-MHz and the 1720-MHz profiles shown in Fig. 9. The 1612-MHz absorption
from $-12$ \kms\ to $-4$ \kms\, followed by emission to $+1$ \kms\, is
matched by a conjugate variation at 1720 MHz. This is a typical example of
the anomaly flip discussed in Section 3.3.1, and the results can be
interpreted in terms of low-density gas for the more negative observed
velocities and high-density for the more positive observed velocities. The
distribution of the absorption and emission at 1612 and 1720 MHz follows
the 1.6-GHz continuum emission (see Fig. 6) and is consistent with an
overlying OH cloud.

\item {\it NGC 6334-F:} Information about the extended OH is very
restricted because the main-line and 1720-MHz observations are dominated by
bright maser emission (see Fig. 7). The 1612-MHz transition is also in
emission centred near the continuum emission peak, at RA(B1950) =
17$^{h}$17$^{m}$32\fs42 $\pm 0.01$, Dec(B1950) =
$-35^{\circ}$44$\arcmin$05\farcs4 $\pm 0.1$. The narrow velocity width of
1.8 \kms\ and the unresolved angular size would be consistent with maser
emission. However the peak intensity of 0.19 Jy beam$^{-1}$ is at least two
orders of magnitude lower than the intensities of the other masers. In
addition, the position differs by 1.8 arcsec from the position of the other
masers and the velocity of the peak ($-8.2$ \kms) is more positive than the
velocities of the 1720-MHz and main-line masers ($-8.9$ and $-10.2$ \kms\
respectively). Interpretation of the results is difficult without
additional information. The presence of several infrared objects and masers
(see Fig. 6) suggest a region of considerable star-forming
activity. Excitation of a high-density OH cloud by infrared radiation could
be responsible for the 1612-MHz emission and possibly the other OH
masers. On the other hand, the intensity of the 1720-MHz and the absence of
1612-MHz absorption may indicate that collisional excitation is
predominant.

\end{description}

\section{Discussion}

\subsection {Overall cloud structure} 

The OH velocity and line-to-continuum trends support a bar-like shape for
the molecular cloud/dust complex in the following manner. The cloud may
wrap behind NGC 6334-C (or disappear) and then curve forward towards NGC
6334-E and NGC 6334-A. This shape was first put forward by
\citeN{Moran80}. They used the sequential theory of \citeN{Elmegreen77} to
suggest that star formation in the molecular cloud may have been triggered
by an OB association located behind it. If the cloud were shaped like a
tube which is bent away from the OB association (resulting in more negative
velocities towards the ends) then a shock wave traveling from the OB
association would hit the central part of the cloud first and then move
outwards towards each end. This model was initially used to explain why the
active sites of star-formation (NGC 6334-V and -F) are located at the end
of the molecular cloud and the more evolved regions (NGC 6334-C and -D)
nearer the centre. However, this model has been complicated by evidence of
ongoing star-formation, such as infrared sources, outflows and OH maser
emission (this paper) towards the central continuum concentrations. Perhaps
a burst of secondary star formation had begun in these regions.

As stated previously, the line-to-continuum ratio is related to OH column
densities but interpretation of the results can be confused if the
continuum emission is located in front of the absorbing gas, which may be
the case for NGC 6334-C and D. In regions where the OH gas appears to be
concentrated near the edges of the continuum emission, the OH may be
gradually becoming dissociated by the expanding ionized regions or used up
for star formation. The column density of the extended molecular cloud
increases towards each end of the continuum axis. This could result from
the increased line-of-sight path-length at the ends of the forward-curving
molecular cloud. Alternatively, it would be consistent with a gas supply
not yet dissipated by existing \HII\ regions or used up in star
formation. If NGC 6334-C were more evolved than the other sources, it is
feasible that the surrounding molecular cloud would be dissociated first.

\subsection{Satellite-line anomalies} 

Several cases of flips across the satellite-line profiles have been
identified and they have been interpreted as a density gradient. There are no
complete models which can explain these flips astrophysically. To explain
their presence in our data, we suggest a simple model whereby an \HII\
concentration is located behind a more extended OH cloud. The OH cloud has
a velocity range of $-15$ to $+2$ \kms\ and consists of a low density
layer, producing 1612-MHz absorption, and an adjacent high density layer,
producing 1612-MHz emission. The transitional density is given by {\it
N}$_{OH} \approx 10^{15} \Delta {\mathrm V}$ km$^{-1}$ s cm$^{-2}$. The
high density layer, with velocities of $-6$ to $+2$ \kms\ and which also
include the \HII\ velocities, is nearer the \HII\ region and the low
density layer, with velocities of $-15$ to $-4$ \kms, is further out along
the line of sight.

The adjacent 1612-MHz emission and absorption observed at the transitional
velocity in NGC 6334-A ($-4$ \kms\ of Fig. 11) may be explained if the
direction of the density gradient within the OH cloud differs from the
direction of the velocity gradient, and that isovelocity planes do not
coincide with isodensity planes. The angle of deviation may be large enough
so that at the transitional velocity it is possible to observe the low
density gas on one side of the continuum emission and the high density gas
on the other.

\section{Conclusion}

We have imaged the 1.6-GHz continuum and the four transitions of the
$^{2}\Pi_{3/2}$, J=3/2 ground state of OH towards the NGC 6334 \HII\
region/molecular cloud complex. Observations were made with the Australia
Telescope Compact Array with angular resolutions of $4.6 \times 7.2$ arcsec$^2$
and $25.8 \times  19.6$ arcsec$^2$. The results are as follows:

\begin{description}
\item (a) {\it 1.6-GHz Continuum distribution} - The continuum emission has
been detected over an area of $35 \times 25$ arcmin$^2$. The main region
shows a set of aligned concentrations in stages of evolution ranging from
compact regions to well-evolved shells. They are superimposed on a faint
extended area. Two additional southern regions appear as extended
incomplete shells.

\item (b) {\it OH maser emission} - We have detected seven masers, plus a
possible faint maser candidate. Including the eighth possible maser, the
masers are present for all OH transitions and appear to be associated with
the continuum concentrations. Two of the masers were detected towards NGC
6334-A. All maser locations include main-line transitions.  The maser
velocities vary between $-8.7$ and 0.5 \kms, with the more positive
velocities located near the centre of the continuum distribution.

\item (c) {\it Extended OH distribution} - Main-line OH absorption was
detected towards almost all of the continuum emission. The distribution of
the line-to-continuum ratio at 1667-MHz ratio is uniform across the
continuum and varies between $-0.4$ and $-0.9$.  The distribution is not
enhanced towards the continuum emission but is concentrated in a bar
extending northeast--southwest with its sharp south-eastern edge coincident
with that of the optical dust lane. Our velocity and line-to-continuum
results support the notion in which the molecular bar wraps around the
continuum emission.

\item(d) {\it Non-LTE behaviour}

For the satellite lines, one transition is commonly in emission with the
other simultaneously in absorption. The 1612-MHz transition is in emission
near the prominent radio continuum concentrations, presumably reflecting OH
excitation by the infrared radiation associated with these
concentrations. In some cases, the conjugate-type behaviour is observed to
reverse across a spectral-line profile. This has been interpreted as a
density gradient, and we have suggested that this reflects the presence of
different cloud regimes along a line of sight. The main-line transitions
show non-LTE behaviour in that the $T_{1667}/T_{1665}$ intensity ratios
range from 1.0 to 1.2, significantly lower than the LTE value of 1.8. The
results of the OH Sum Rule suggest that this discrepancy can only be
explained in terms of high optical depth if the cloud is clumping on
smaller scales.

\end{description}

\section*{Acknowledgments}

We thank Neil Killeen and Bob Sault for assistance with the data reduction
and analysis and John Storey and Jochen Liske for valuable discussions. We
also thank Jim Caswell and the referee for providing some very helpful
comments. KJB acknowledges the support of an Australian Post-graduate
Award.

\end{document}